\documentclass[manuscript,nonacm]{acmart}
\pdfoutput=1
\usepackage{amsmath,graphicx}
\usepackage{multirow}
\usepackage{url}
\usepackage{hyperref}
\usepackage{xcolor,colortbl}
\usepackage{makecell}
\newcommand{\RQ}[2]{\label{sec:rq#1} \newline \noindent \emph{\textbf{RQ#1}~$\cdot$~#2} \newline}

\AtBeginDocument{%
  }

\begin{document}

\title{Every Breath You Don't Take: Deepfake Speech Detection Using Breath}

\author{Seth Layton}
\email{sethlayton@ufl.edu}
\author{Thiago De Andrade}
\email{tdeandradebezerr@ufl.edu}
\author{Daniel Olszewski}
\email{dolszewski@ufl.edu}
\author{Kevin Warren}
\email{kwarren9413@ufl.edu}
\author{Kevin Butler}
\email{butler@ufl.edu}
\author{Patrick Traynor}
\email{traynor@ufl.edu}
\affiliation{%
  \institution{University of Florida}
  \streetaddress{}
  \city{Gainesville}
  \state{Florida}
  \country{USA}
  \postcode{}
}



\renewcommand{\shortauthors}{Layton et al.}

\begin{abstract}
  Deepfake speech represents a real and growing threat to systems and society. Many detectors have been created to aid in defense against speech deepfakes. While these detectors implement myriad methodologies, many rely on low-level fragments of the speech generation process. We hypothesize that breath, a higher-level part of speech, is a key component of natural speech and thus improper generation in deepfake speech is a performant discriminator. To evaluate this, we create a breath detector and leverage this against a custom dataset of online news article audio to discriminate between real/deepfake speech. Additionally, we make this custom dataset publicly available to facilitate comparison for future work. Applying our simple breath detector as a deepfake speech discriminator on in-the-wild samples allows for accurate classification (perfect 1.0 AUPRC and 0.0 EER on test data) across 33.6 hours of audio. We compare our model with the state-of-the-art SSL-wav2vec model and show that this complex deep learning model completely fails to classify the same in-the-wild samples (0.72 AUPRC and 0.99 EER).
\end{abstract}



\keywords{breath, synthetic, speech, audio, detection, deepfake}


\maketitle

\section{Introduction}
\label{sec:intro}
Deepfake speech (e.g., text-to-speech, deepfakes, voice assistants) aims to make the differentiation between synthetic and organic speech difficult~\cite{lorenzo-trueba_can_2018, SaundersDetectingDF}. While such audio has many benign uses, the potential for dangerous applications has created the need for accurate and automated demarcation of human-spoken from synthetically-generated audio.

The research community has responded with competitions such as ASVspoof~\cite{wu2015asvspoof, kinnunen2017asvspoof, todisco2019asvspoof,yamagishi2021asvspoof}, ADD~\cite{yi2022add}, and SASV~\cite{jung2022sasv}. These competitions curate datasets of deepfake and real speech and invite participants to create detection algorithms to test on these datasets. Subsequently, these datasets are the de facto standard for deepfake speech and give a baseline of comparison for all current and future deepfake speech detectors. Most of the currently existing speech deepfakes detectors focus on low-level spectral (e.g., spectrogram, MFCC, LFCC, and CQCC) imperfections created during the audio generation pipeline and show starkly different classification results vs. human interpreters~\cite{WTC_24}. This technique of low-level spectral detection will be rendered obsolete due to the rapid advancement of the speech generation field. Thus, a paradigm shift towards high-level speech features such as prosody detection~\cite{attorresi2022combining}, emotion detection~\cite{hosler2021deepfakes}, and anatomical shape detection~\cite{blue2022you} is underway.

One promising avenue for high-level speech feature exploration is breath, as breathing is one of the subtle ways that humans subconsciously perceive naturalness in speech~\cite{braunschweiler_automatic_2013, elmers_take_2021, szekely_breathing_2020}. Additionally, the demand for automatic breath detection methods for medical research purposes is elevated due to COVID-19~\cite{deshpande_overview_2020}. This demand influenced the INTERSPEECH 2020 Computational Paralinguistics Challenge to create the Breathing Sub-Challenge, a track dedicated to breath detection~\cite{schuller_interspeech_2020}. Current state-of-the-art breath detection methods include spectral-based models~\cite{hamke_breath_nodate, nallanthighal_deep_2021}, acoustic models~\cite{braunschweiler_automatic_2013}, and raw speech waveform deep learning~\cite{nallanthighal_deep_2021}. While these methods automatically detect breath in audio, applying these techniques to the realm of deepfake speech to demarcate real from speech deepfakes is an open challenge. Towards this, we explore the viability of breaths as a detection mechanism for deepfake speech. 

Our work focuses on in-the-wild deepfake speech. News outlets inherently opt for industry-standard high-quality synthetic speech to not deter listeners. We note that there is a distinction between industry-standard and state-of-the-art. This distinction creates a mismatch between academically-created and in-the-wild deepfakes. Specifically, academia produces cutting-edge deepfakes, but these deepfakes are rarely seen en masse on the internet. It is thus important to study what is currently affecting the largest population and is the focus of this work. Towards this, we gather speech samples from online news vendors that employ a text-to-speech or human-read audio listening option for online text articles. We then determine the efficacy of breath detection as a deepfake speech discriminator using simple models. We compare our simple models against a state-of-the-art model and highlight the shortcomings of complex machine learning techniques. We focus on currently deployed deepfake speech, which gives a measure for the industry standard in the field and as such, we argue against the use of popular academic datasets such as ASVspoof. 

Our main contributions are: 
\begin{itemize}
  \item We perform a study of currently deployed in-the-wild synthetic speech.
  \item We create a generation-agnostic deepfake speech detector based solely on breaths.
  \item We publish a dataset of in-the-wild text-to-speech and human-read speech.
  \item We highlight shortcomings and over-reliance on complex deep learning detection models.
\end{itemize}

The remainder of this paper is organized as follows:
Section~\ref{sec:hypothesis} states our hypothesis;
Section~\ref{sec:meth} details our methodology;
Section~\ref{sec:experiments} presents our experimental results;
Section~\ref{sec:disc} offers discussion and insights based on our findings;
Section~\ref{sec:rel_work} discusses related work; and
Section~\ref{sec:conc} presents our concluding remarks.

\section{Hypothesis}
\label{sec:hypothesis}

We hypothesize that current speech deepfakes generation techniques do \emph{not} sufficiently incorporate breaths.

To investigate this claim, we must first determine if breath is a generalizable speech feature. Thus, we define our first research question:
\RQ{1}{Are breaths automatically detectable, intra and inter-speaker?}

As breath is one of the ways that humans determine naturalness in speech; we define our second research question:
\RQ{2}{Do current deepfake voices generate breaths?}

Combining the previous research questions we define our final research question:
\RQ{3}{Are breaths able to accurately discriminate between in-the-wild deepfake and real samples?}

\section{Methodology}
    To determine if breaths are a generalizable feature between speakers and subsequently useful as a discriminator against real and fake speech we define deepfake speech, gather real and fake data, implement and test detection models, and evaluate the performance of breathing.

\label{sec:meth}
    \subsection{Cheapfake vs. Deepfake}
    \label{sec:meth:cvd}
        Synthetic media is a spectrum that spans from generative (e.g., machine learning and artificial intelligence) to manually altered (e.g., Photoshop and speech waveform manipulation) samples. The use of deep neural networks creates a deepfake and using cheap manual software manipulation creates a cheapfake~\cite{paris2019deepfakes}. Specifically, a cheapfake requires a pre-existing sample of a specific individual for manual modification; whereas, a deepfake may be fully generative and not require a source sample. While both forms of synthetic media influence society, we focus on the dominant form of synthetic media known as deepfakes for the remainder of this paper. More critically, we only consider media samples that are entirely real or entirely fake (i.e., no partially fake samples with segments altered). Additionally, this extends to manipulation techniques such as altering the rate of speech or changing the pitch.

    \subsection{Dataset}
    \label{sec:meth:dataset}
        We employ a multi-tiered model pipeline that requires independent datasets during the training phase. We gather single-speaker podcast audio for training the breath detector and online news articles read by humans and text-to-speech algorithms for the final deepfake speech detection training and testing.

        \subsubsection{Why not ASVspoof} 
        \label{sec:meth:dataset:no_asvspoof}
        \hfill\\
            ASVspoof is the de facto standard dataset when creating and testing a deepfake speech detector and it gives a community baseline for comparison. However, for our task, this dataset is not representative, sufficient, or realistic. First, our model relies on features of breathing and thus samples must be sufficiently long to contain a breath (\emph{93\%} of ASVspoof 2021 samples are shorter than 5 seconds). As breaths in read and spontaneous speech are expected at a rate of 8-14 per minute~\cite{nallanthighal2019deep}, each ASVspoof sample is not expected to contain any of these important high-level features. Additionally, many of the samples longer than 5 seconds are filled with incoherent speech due to generation issues. Moreover, the class distribution of the ASVspoof 2021 samples is unrealistic (\emph{97\%} deepfake and 3\% real in the evaluation set). This class distribution is vastly different from the expected in the wild distribution and would require multiple resampling techniques to correct, which could bias the results due to the vast imbalance of the dataset~\cite{needlestack_layton}. Finally, we opt to gather samples that are \emph{currently} in the wild today as a better representation of deepfake speech and contain a balanced real/fake class distribution.

        \subsubsection{Podcasts}
        \label{sec:meth:dataset:podcasts}
        \hfill\\
            We curate a training dataset of podcasts that meet specific criteria and manually annotate these podcasts for breath locations. Each podcast is single-speaker, contains no background music, is free from obvious background noise, and breathing is noticeable. In total, we selected 10 podcasts from 4 speakers totaling just over 5 hours. Each podcast is manually annotated for precise breath locations by two independent annotators and a third annotator verifies and reconciles any discrepancies in the annotations from each initial annotator to ensure all breaths are captured.
            \hfill\\

        \subsubsection{News Articles}
        \label{sec:meth:dataset:news_articles}
        \hfill\\
            We gather news article audio from online news vendors for training and testing our deepfake speech detector. We search for news articles with strings like ``Listen to this article'' and  ``Hear the full article'' keywords on the webpage to indicate that there is an audio version of the transcript available. Manual checking of the news vendor is required to determine if the news articles provided are text-to-speech generated, or spoken by a human. In total, we collected 277 TTS (26.94 hours) and 56 human-read (25.54 hours) news articles from four different news outlets (two TTS sets and two human-read sets).\footnote{Each sample is entirely deepfake, or entirely real. There is no scenario for deepfake speech injected into real audio. We do not redistribute the audio; however, links to news article websites are freely available at: \url{https://sites.google.com/view/ebydt/}.}

    \subsection{Breath Detection}
    \label{sec:meth:breath_detect}
        First, we create a breath location detector to highlight breaths in a given audio sample. Towards this, we take inspiration from rare-event detection~\cite{szekely2019casting, amiriparian2018deep} to create our pipeline for breath detection. Unlike these works, however, we do not perform image analysis on spectrograms, but instead use the raw values computed from the spectrogram. Thus, all the layers of our model lose a dimension, reducing model complexity and providing speed increases to training and inference which helps real-time predictions.

        \begin{figure}[t]
            \centering
            \includegraphics[width=\columnwidth]{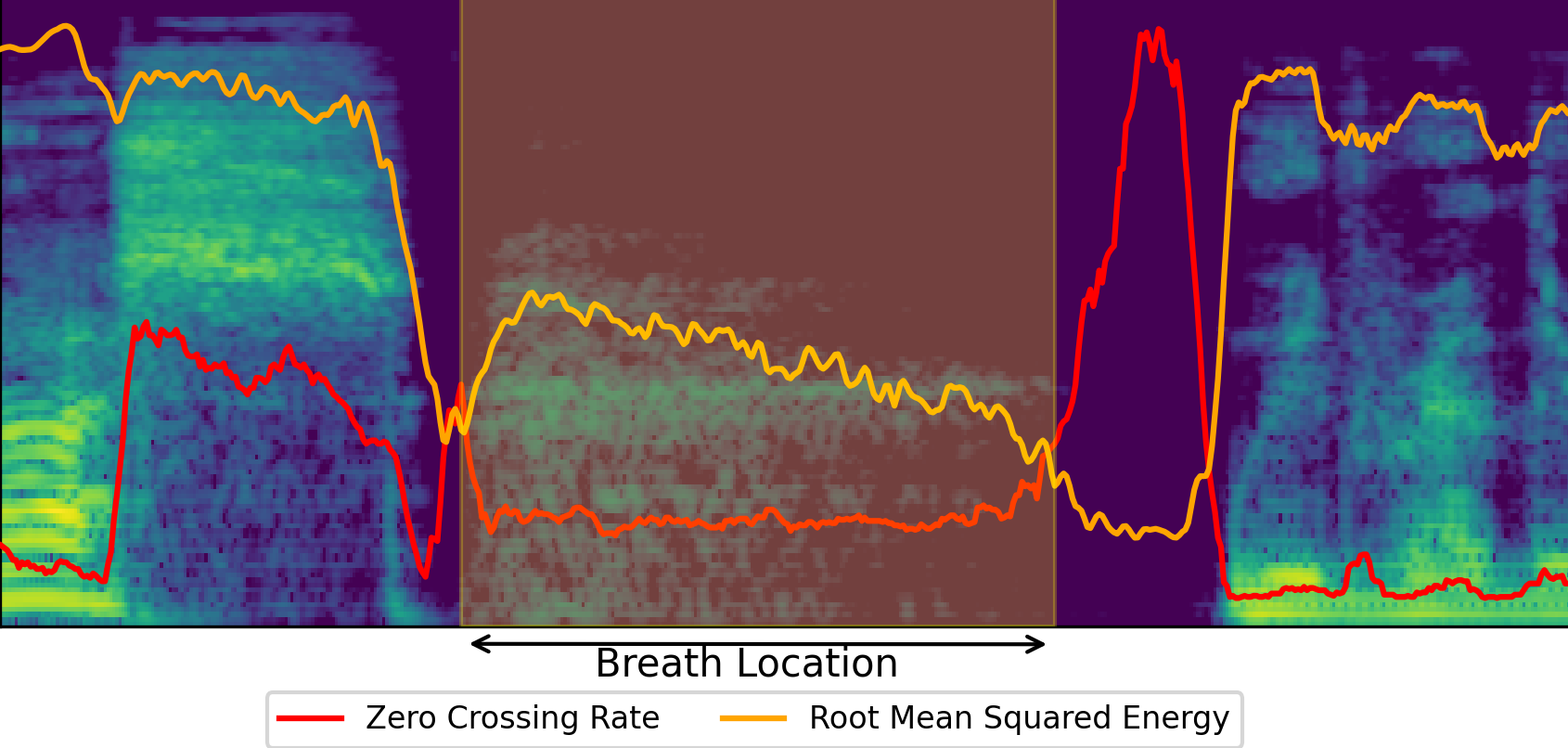}
            \caption{A visual representation for a segment of speech containing a breath using a $window\_length$ of 20ms and a $hop\_length$ of 2.5ms. During the spoken segments before and after the breath RMSE is at peak values while the ZCR is at minimum values. Immediately surrounding a breath is a non-voiced segment where the RMSE values drop and ZCR values rise, but then both move to a medium value during the breath. Additionally, the background mel spectrogram shows higher energy across all frequencies during spoken segments, medium energy at lower frequencies during breaths, and relatively little energy at all frequencies for silence.}
            \label{fig:bubble}
          \end{figure} 

        \subsubsection{Feature Selection} 
        \label{sec:meth:breath_detect:feats}
        \hfill\\
            We first calculate raw values of the mel-spectrogram (dB converted), zero crossing rate (ZCR), and root mean squared energy (RMSE) (dB converted). Each of these features slides a window over a waveform to calculate an overlapping temporal-based value. As the size of the window and the duration between windows may be optimized we test a variety of values ranging from 5ms -- 200ms $window\_length$ and 2.5ms -- 25ms $hop\_length$. The result of this feature extraction is an array of frames that contain spectrogram, ZCR, and RMSE values for an entire audio sample. The number of frames in an array is calculated as ${num_{frames} = \frac{sample\_duration(milliseconds)}{hop\_length}}$, and the value of each frame is the aggregated mel-spectrogram, ZCR, and RMSE for $window\_length$ duration. For example, a 5-second excerpt of audio with a $window\_length$ of 50ms and a $hop\_length$ of 5ms creates a 1000-frame array.  We test a spectrum of sizes, shapes, and durations for these features and select a $window\_length$ of 20ms, a $hop\_length$ of 2.5ms, and 128 mel-spectrogram buckets for our final model as these produce the best results.

            Next, using the manually annotated location of breath we denote each frame in the array as either a breath (i.e., positive class) or not a breath (i.e., negative class). A frame is considered to contain breath if more than half the $window\_length$ of that frame is annotated as a breath. Figure~\ref{fig:bubble} shows our selected features and how they change during a breath and speaking before/after breathing. For breaths, ZCR and RMSE tend towards medium values between silence and spoken segments and the mel-spectrogram shows only energy at lower frequencies.

        \subsubsection{Model Architecture} 
        \label{sec:meth:breath_detect:model}
\begin{figure}[t]
        \centering
        \includegraphics[width=\columnwidth]{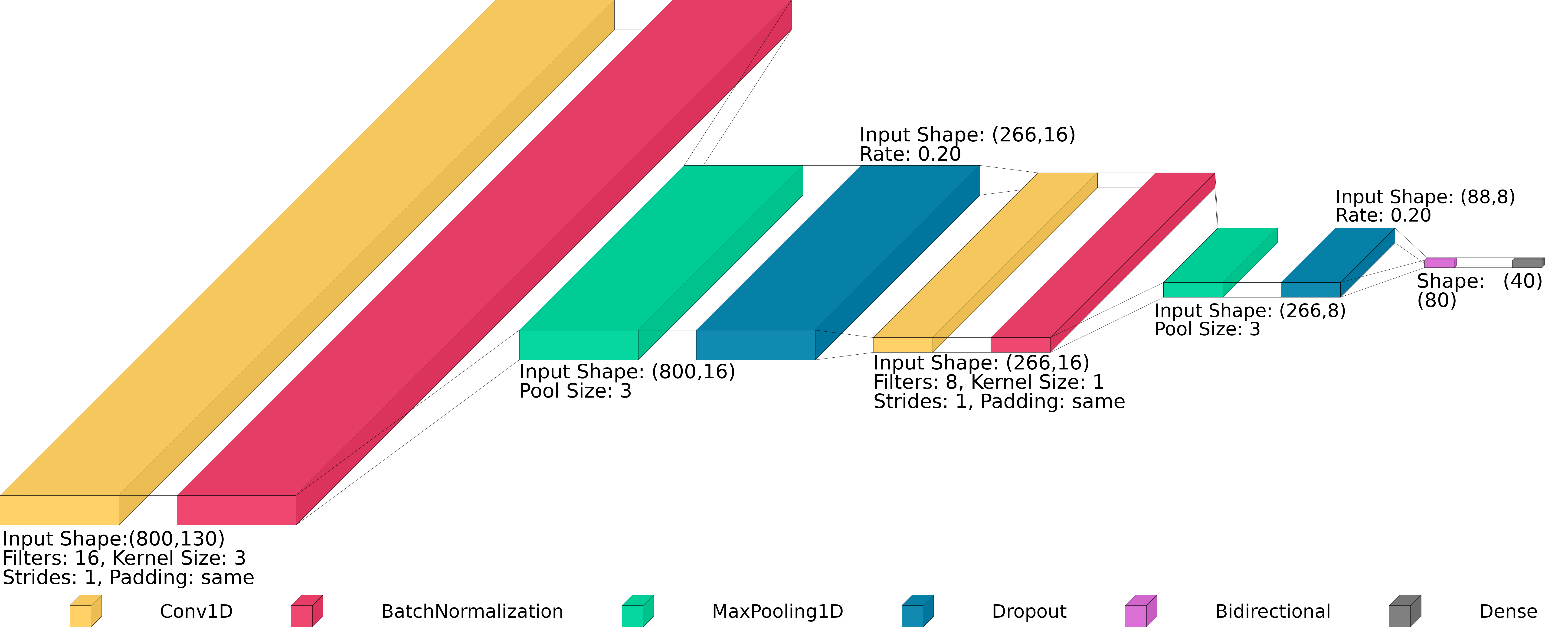}
        \caption{A visual representation of the breath detection model architecture.}
        \label{fig:pipeline:a}
    \end{figure}

        \hfill\\
            Based on Székely et al.~\cite{szekely2019casting} we build a multi-tiered convolutional and recurrent neural net detection model. The model architecture starts with two 1D convolutional layers (16 and 8 filters, 3 and 1 kernel sizes, same padding, and ReLU activation); each convolutional layer is followed by batch normalization, max pooling (pool size of 3), and 0.2 dropout. These initial layers are used as input to a bidirectional-LSTM layer which feeds into a dense layer with a sigmoid activation for final prediction. We use a binary cross-entropy loss as the loss function and Adam as the optimizer function. The pipeline for this model is shown in Figure~\ref{fig:pipeline:a} (using a $window\_length$ of 20ms, a $hop\_length$ of 2.5ms, 128 mel-spectrogram buckets, and a batch size of 32). We experimented with many different shapes and sizes for our breath detection model architecture such as changing the total number of convolutional layers, adjusting the number of filters and kernel sizes of each convolutional layer, altering the default max pool size, and changing the size of the dropout layer. Ultimately we select the sizes and parameters defined in Figure~\ref{fig:pipeline:a}.

            As input to our breath detection model, an entire audio sample is sectioned off into 2-second segments and sequentially fed to the model with predictions given for every 50ms (i.e., 40 predictions per 2-second chunk). We opt for 2-second slices of audio as increasing the duration of the slices subsequently increased training and inference time without improving performance. A 2-second segment is 800 2.5ms slices of features and there are 128 values from the mel-spectrogram and 1 feature from each of the ZCR and RMSE giving a total of 130 features. The shape of (32 x 800 x 130) is given as input to our breath detector as we use a batch size of 32. The output for a 2-second chunk of audio is 40 sequential binary classifications as to whether or not there is a breath located in each 50ms slice.

        \subsubsection{Model Result Post-Processing} 
        \label{sec:meth:breath_detect:model_pp}
        \hfill\\
            Our pipeline implements a simple mechanism on the resultant predictions. We remove any predicted breaths that are shorter than 150ms as we measured no breaths shorter than 150ms in any of our podcasts. This post-processing affects few samples, yet helps smooth out the resultant breath locations. We then use these breath locations as input features for the deepfake speech detection algorithm.  
        
        \subsubsection{Metrics}
        \label{sec:meth:breath_detect:metrics}
            We use a multitude of metrics for our full pipeline and evaluation of intermediary and final results. For breath detection, we use the binary cross entropy loss function for hyperparameter tuning and use Area Under the Precision-Recall Curve (AUPRC) for evaluating model performance. We use AUPRC as our main metric as it is robust to imbalanced data~\cite{arp2022and, fu2019tuning} and effectively highlights performance on the important class. As breaths are an imbalanced class distribution problem (i.e., non-breathing heavily outweighs breathing in normal speech) and are the important class, AUPRC is a suitable metric for honestly evaluating performance. 

    \subsection{Deepfake Speech Detection}
    \label{sec:meth:df_detect}
\begin{figure}
    \centering
    \includegraphics[width=\columnwidth]{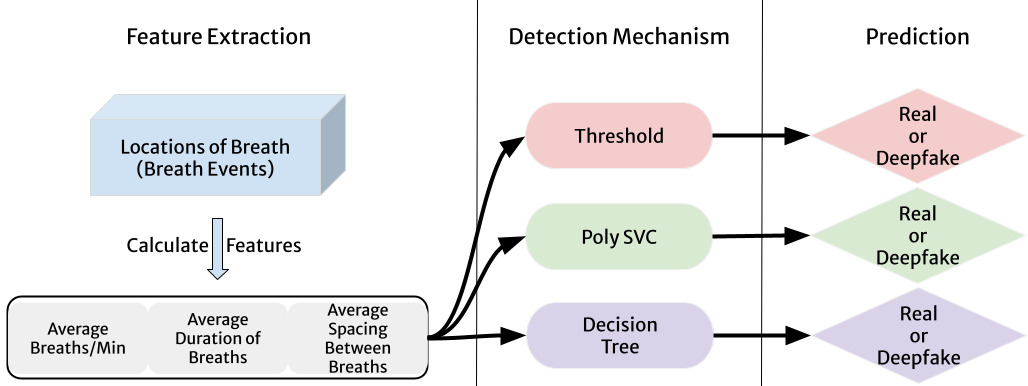}
    \caption{A visual representation of the final stage of the detection pipeline. We use/compare three different simple classifiers in the last stage to showcase the relative interchangeability of models for final prediction.}
    \label{fig:pipeline:b}
\end{figure}

        The final component of our pipeline uses the predicted breath locations as input. For an audio sample, we use the predicted breath locations to calculate three features: average breaths per minute, average breath duration, and average spacing between breaths. With these features, we create multiple simple deepfake speech detectors to test the viability of computationally inexpensive methods and compare models to a complex deep learning model (Section~\ref{sec:experiments:df_detect:comp}). The first is a thresholding classifier that uses previously measured breath statistics. The second is a C-Support Vector Classification (SVC) algorithm with a poly kernel, a regularization parameter of 1, and a degree of 2. Finally, we implement a three-tiered decision tree with default parameters. Each of these mechanisms produces a single binary prediction for the entire audio sample. The pipeline for this deepfake detection model is shown in Figure~\ref{fig:pipeline:b}. 

        The thresholding method should be sufficient as a final discriminator if deepfake speech does not produce any breaths. However, if breaths are produced, then a more robust mechanism would be required to accurately discriminate between real and fake samples. Thus, we apply both techniques and compare the results to determine viability and then compare them to complex and computationally expensive state-of-the-art.

        \subsubsection{Metrics}
        \label{sec:meth:df_detect:metrics}
           For the final deepfake speech detector, we implement a range of metrics to help contextualize model performance, as single metrics may fail to honestly describe model performance. These metrics are accuracy, F1-score, precision, recall, AUPRC, equal error rate (EER), true positives, true negatives, false positives, and false negatives. For the thresholding method, we are unable to use AUPRC and EER as these metrics require probabilities, and only a prediction is generated. Finally, we note that the use of EER as a performance metric is deprecated by ISO/IEC standards~\cite{evalplan2021asvspoof} and has been shown to obfuscate results~\cite{brummer2021out, sugrim2019robust}; however, EER remains the current community standard for deepfake speech detection.

\section{Experiments}
\label{sec:experiments}    
    To sufficiently answer our research questions defined in Section~\ref{sec:hypothesis} we perform two experiments starting with breath generalizability and finishing with deepfake speech (deepfake) detection.
    \subsection{Breath Generalizability (\textbf{RQ1})}
    \label{sec:experiments:breath_generalize}
        To understand if breath sounds are generalizable between different individuals (thereby justifying them as a deepfake speech discriminator), we create three tests. RQ1 tests the generalizability of breathing with Test 1 obtaining a comparative baseline for successive tests, Test 2 examining the performance of each podcast, and Test 3 examining the performance of each speaker. Each test uses our breath detection model defined in Section~\ref{sec:meth:breath_detect} without any changes.

      \subsubsection{Test 1} 
      \label{sec:experiments:breath_generalize:test1}
      \hfill\\
        Test 1 creates a best-case baseline for comparison with Tests 2 and 3 employing k-fold cross-validation. Test 1 takes $(\frac{1}{x} * 100)\%$ consecutive frames, where $x$ is the total number of podcasts, randomly from each podcast to use as the validation set and uses the remaining $100-(\frac{1}{x} * 100)\%$ from each podcast as the training set. We do this process 100 times and calculate the validation AUPRC to obtain a baseline for comparison. This test gives the best-case scenario due to having a subsample of every podcast/speaker in each training and validation batch. This is in contrast to the following two tests which avoid overlapping training and validation speakers/podcasts.

      \subsubsection{Test 2}
      \label{sec:experiments:breath_generalize:test2}
      \hfill\\
        Test 2 employs a leave-one-out strategy to determine the impact on breath detection for each podcast. A podcast is set aside individually as the validation set and the remaining $x-1$ podcasts become the training set. We retrain our breath detection model using the training set and calculate the AUPRC for the validation set. We do this for all $x$ podcasts and compare the results to Test 1 to highlight any podcasts that may be problematic, or not generalizable.

      \subsubsection{Test 3}
      \label{sec:experiments:breath_generalize:test3}
      \hfill\\
        While similar to Test 2, Test 3 evaluates the impact on breath detection for a single speaker. Towards this, all podcasts from a specific speaker are set aside as the validation set and the remaining speaker's podcasts become the training set. Again we retrain our breath detection model using the training set and calculate the AUPRC for the validation set. This process is repeated for each speaker and compared against Test 1 to determine if breaths are generalizable between speakers.
      \\
      \begin{figure}
        \begin{minipage}[b]{1.0\linewidth}
          \centering
          \includegraphics[width=\columnwidth]{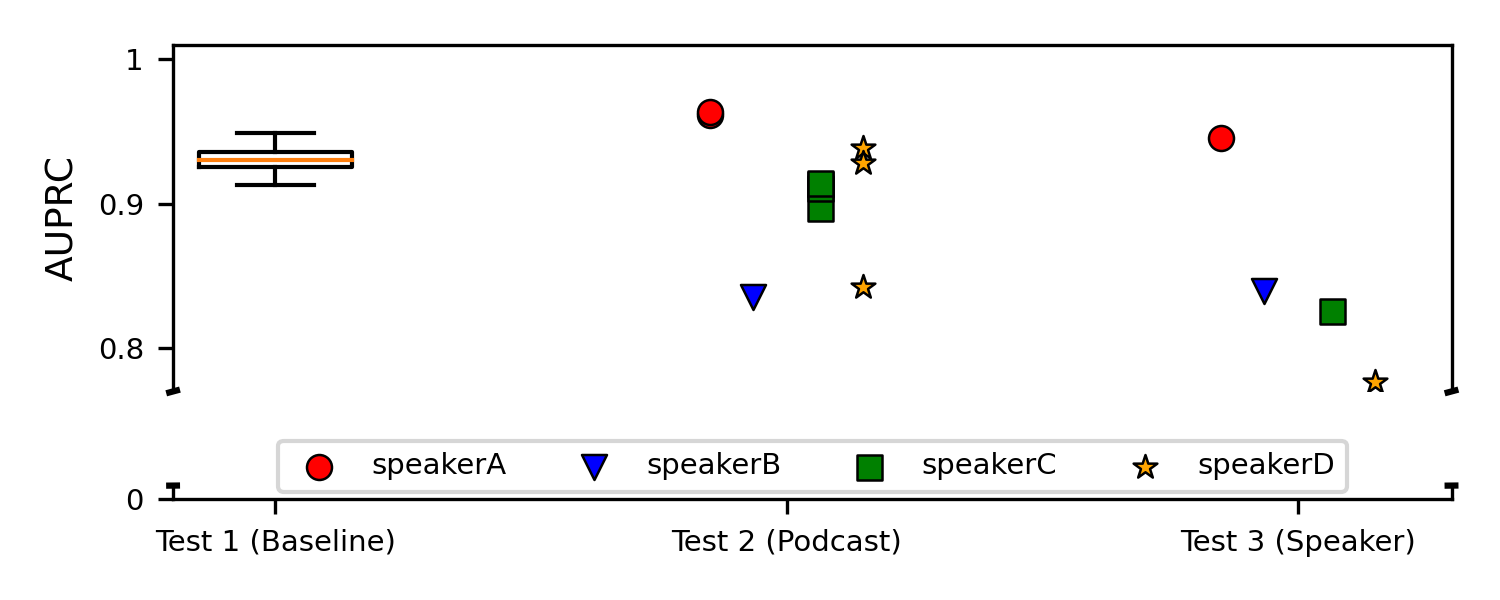}
        \end{minipage}
        \caption{The baseline validation testing on all podcasts vs. the leave-one-out testing for each podcast and each speaker. Each point is a specific speaker/podcast as the validation set. We show that breath measured in this capacity is generalizable and thus useful as a deepfake speech discriminator.}
        \label{fig:podcastfeats}
      \end{figure}

      \subsubsection{Results}
      \label{sec:experiments:breath_generalize:results}
      \hfill\\
        Figure~\ref{fig:podcastfeats} shows the results of these three tests. We use the baseline AUPRC of $\sim$0.93 from Test 1 to compare Tests 2 and 3. Test 2 shows podcast-specific leave-one-out performance between $\sim$0.84 and $\sim$0.95, which is a marked reduction in performance from the baseline (Test 1). Test 3 shows speaker-specific leave-one-out performance between $\sim$0.78 and $\sim$0.94, which is also a reduction in performance from the baseline. This reduction is expected as in Test 1 there is training data of similar distribution to the validation in every iteration since validation is derived as a subset of each podcast as described in Section~\ref{sec:meth:breath_detect}. However, we see an AUPRC mean and standard deviation of 0.931$\pm$0.008, 0.911$\pm$0.041, and 0.84$\pm$0.062 for Test 1/2/3 respectively. This indicates that breaths are generalizable and thus usable as a general discriminator for deepfake speech (\textbf{RQ1}).

    \subsection{Deepfake Speech Detection (\textbf{RQ2}, \textbf{RQ3})}
    \label{sec:experiments:df_detect}

        \subsubsection{Setup}
        \label{sec:experiments:df_detect:setup}
\begin{figure}
        \centering
        \includegraphics[width=\columnwidth]{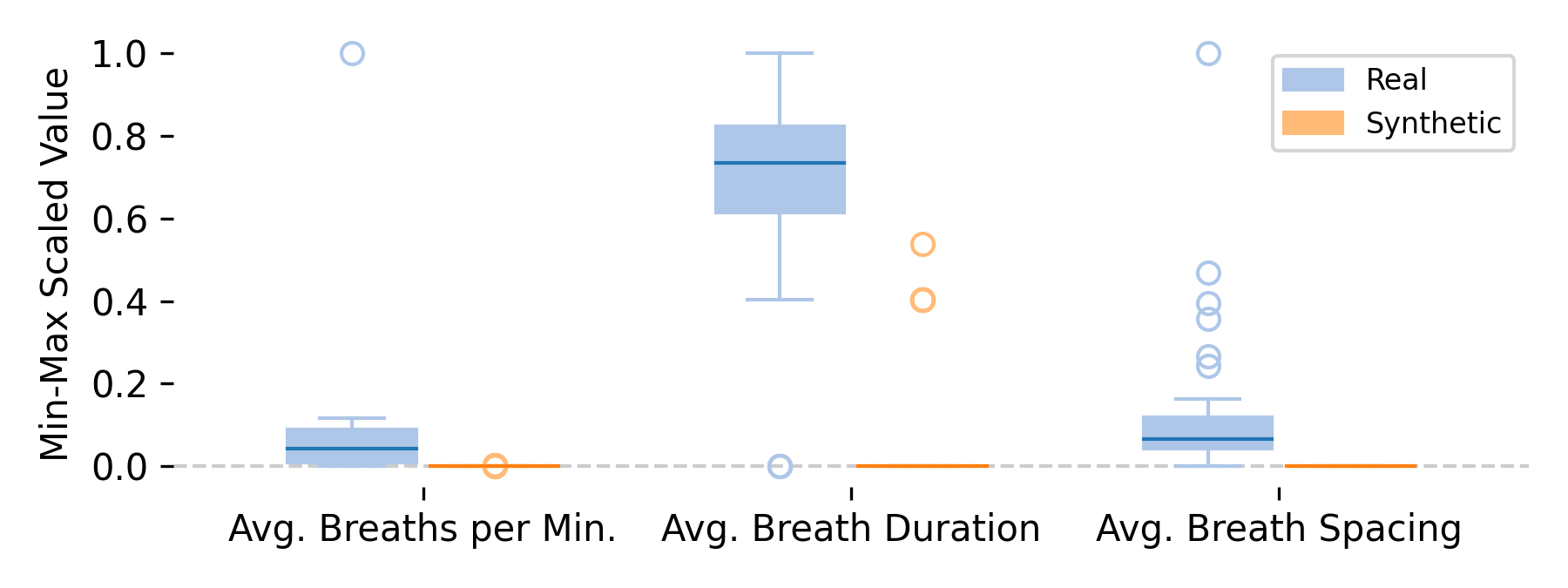}
        \caption{We show that there is a clear distinction (i.e., virtually no overlap) between human-read and synthetically-generated news articles with respect to breath statistics. The only overlap present is in outliers from each type of speech.}
        \label{fig:newsartfeats}
    \end{figure}

        \hfill\\
            Now that we have shown the generalizability of breaths between individuals, we apply this methodology as a discriminator for deepfake and real speech using the three methods described in Section~\ref{sec:meth:df_detect}. First, we train our breath detection model on all the podcast data to ensure the best possible final model. Next, we pass the news article data outlined in Section~\ref{sec:meth:dataset:news_articles} to this final model to get breath locations for each news article. We then use the breath locations and calculate the three features detailed in Section~\ref{sec:meth:dataset}. Figure~\ref{fig:newsartfeats} shows the clear delineation between the deefpake and real news article breath features. We demonstrate that deepfake speech, as seen in the wild today, \emph{does not appropriately produce breaths}, which indicates a strong discriminator using these features (\textbf{RQ2}). Finally, we split the news articles into a training and testing set where no samples from the same news outlet are in the training and test sets. This split comes out to 101 training samples (18.9 hours) and 232 testing samples (33.6 hours) and ensures no bias when testing.
            
            For the poly SVC and decision tree methodologies, we train our model (Section~\ref{sec:meth:df_detect}) with the training set and evaluate performance on the test set. Whereas for the thresholding methodology, we do not need to split the news articles into a training and test set as there is no training taking place since this is not a machine learning model; however, we show results for both the training and test sets to compare against the poly SVC method. Observing Figure~\ref{fig:newsartfeats} we show that in each calculated metric real samples have a non-zero value. We use this as our guideline and set the threshold for each metric to be $Value > 0.0 = Real, Value <= 0.0 = Fake$. For this simple thresholding, we require that all features (i.e., avg. bpm, avg. breath spacing, and avg. breath duration) be $ > 0.0$ to be considered real. Stated concisely, for any sample feature that contains a zero value that sample is considered fake.
        
          \subsubsection{Results}
          \label{sec:experiments:df_detect:results}
          \hfill\\
            Table~\ref{tab:results} shows the training and test results for the news article dataset. We show that a simple SVC model can accurately discriminate between real/deepfake speech samples without overfitting the training set. We show substantial performance across all metrics, and most importantly, \emph{perfect} performance against the test set. Using breaths, we correctly predict all deepfake samples without falsely predicting any real samples (\textbf{RQ3}). Additionally, we show performant results on a varying distribution of data, as the generation techniques for each of the news articles are unknown. Additionally, Table~\ref{tab:results} contains the results for our thresholding methodology. We only display results for the testing set as there is no training required by this technique. Through this, we show that simple thresholding obtains performant results against all news articles. While these results are slightly worse than the results of the SVC model, they do not require any training or testing differentiation and as such, no training time. We posit that simple thresholding is sufficient to capture the nuance in breathing for real and synthetically generated audio samples (\textbf{RQ3}).
      \begin{table}[]
          \renewcommand{\arraystretch}{2} 
          \centering
          \begin{tabular}{cccccccccccccc}
        \cline{2-13}
           &\textbf{Model} & \textbf{Dataset} & \textbf{EER} & \textbf{AUPRC} & \textbf{Accuracy} & \textbf{F1} & \textbf{Precision} & \textbf{Recall} & \textbf{TP}   & \textbf{FP}  & \textbf{TN}   & \textbf{FN} \\ \hline
            \multirow{6}{*}{\rotatebox[origin=c]{90}{\textbf{Simple Models}}} &   $Poly\:SVC$   & Train      & 0.14  & 0.90  & 0.92  & 0.94    & 0.95      & 0.91 & 71   & 5   & 24   & 1      \\
                     &   $(Our Model)$    & Test       & 0.00  & 1.00  & 1.00 & 1.00    & 1.00      & 1.00 & \cellcolor{blue!25}205  & \cellcolor{blue!25}0   & \cellcolor{blue!25}27   & \cellcolor{blue!25}0      \\ \cline{2-13}
                     &   $Decision\:Tree$   & Train      & 0.11  & 0.95  & 0.92 & 0.94    & 0.96      & 0.90 & 72   & 6   & 23   & 0      \\
                     &   $(Our Model)$      & Test       & 0.02  & 0.98  & 0.98 & 0.99    & 1.00      & 0.96 & \cellcolor{blue!25}205  & \cellcolor{blue!25}2   & \cellcolor{blue!25}25   & \cellcolor{blue!25}0      \\ \cline{2-13}
                     &   $Thresholding$         & Train  & -- & --  & -- & -- & --  & -- & --   & --   & --   &  --  \\
                     &   $(Our Model)$                  & Test   & -- & --  & 0.90 & 0.91 & 0.95  & 0.90 & \cellcolor{blue!25}182  & \cellcolor{blue!25}0   & \cellcolor{blue!25}32   &  \cellcolor{blue!25}18  \\ \Xhline{4\arrayrulewidth}

            \multirow{4}{*}{\rotatebox[origin=c]{90}{\textbf{Complex Models}}} & $wav2vec-_{P}$ & Train       & --  & --  & -- & --  & --      & -- & --  & --  & --    & --      \\ 
                      &  ($Pretrained$)                  & Test       & 0.99  & 0.72  & 0.57 &  0.9  & 0.95      & 0.56 & \cellcolor{blue!25}205  & \cellcolor{blue!25}24  & \cellcolor{blue!25}3    & \cellcolor{blue!25}0      \\ \cline{2-13}
                       & $wav2vec-_{R}$                 & Train      & 0.00  & 0.95  & 1.00  & 1.00   & 1.00      & 1.00 & 72   & 0   & 29   & 0      \\
                       & ($Retrained$)                               & Test       & 0.77  & 0.06  & 0.10  & 0.12   & 0.06      & 0.50 & \cellcolor{blue!25}0    & \cellcolor{blue!25}0   & \cellcolor{blue!25}27   & \cellcolor{blue!25}205    \\
          \end{tabular}
          \caption[labelformat=empty]{Results for all simple detection models (poly SVC, decision tree, and thresholding). With the poly SVC model, we obtain performant results on all metrics for the training set, accompanied by a perfect performance on the test set. We demonstrate that the pretrained wav2vec model falls short compared to our model in every metric and augmenting the train set with ``in-distribution'' data and retraining does not solve these shortcomings.}
          \label{tab:results}
        \end{table}

        \subsubsection{Experimental Comparison to Other Detectors}
        \label{sec:experiments:df_detect:comp}
            To fully contextualize the performance of our breath detector we compare our models against a highly complex and heavily trained model: SSL-wav2vec2.0. Towards this, we implement the XLS-R-based deepfake detector~\cite{tak2022automatic}, as this model is the current best performer on the ASVspoof 2021 dataset by a substantial margin. While we do not use the ASVspoof 2021 dataset in any capacity, the XLS-R model was pretrained on 436,000 hours of non-ASVspoof speech data and is sufficiently performant on non-ASVspoof datasets. 
            
            We deploy the pretrained ($wav2vec-_{P}$) model and additionally, to ensure a fair comparison, we fine-tune SSL-wav2vec2.0 with additional training ($wav2vec-_{R}$). We implement this fine-tuning by augmenting the SSL-wav2vec2.0 training data with the 18.9 hours of news articles from our training set. This is the same methodology used to retrain SSL-wav2vec2.0 for ASVSpoof 2021. 
            
            We pass the test set of news articles to both versions of SSL-wav2vec (in 4-second chunks, due to model requirements) and use a probability soft voting scheme to get a single prediction for each audio file. Table~\ref{tab:results} shows the results of these models and shows that our model outperforms in every metric. The pretrained SSL-wav2vec predicts nearly the entire test set as deepfake speech whereas the retrained model predicts all the test data as real speech (while predicting the train set with 100\% accuracy). The highlighted sections in Table~\ref{tab:results} show the classification decision for all news articles in the test set for each of the model types (i.e., SVC, Decision Tree, SSL-wav2vec2.0). We note that $wav2vec-_{R}$ predicts all news articles as real and may seem as though the training phase was incorrectly handled; however, looking at the results of $wav2vec-_{R}$ on the training data shows that it indeed picks up the signal in the data. Through this, we show that our models perform substantially better in every category than the complex models, specifically looking at the false positives it is clear to see that alarm fatigue~\cite{cvach2012monitor} would overwhelm $wav2vec-_{P}$. Comparing this to our models which have essentially no false positives, it is clear to see the potential shortcomings of complex models.

            As the thesis of our paper argues, we demonstrate that models focused on low-level speech features can completely fail when tasked with predicting new data as seen by the contradicting results between $wav2vec-_{P}$ and $wav2vec-_{R}$. Simply put, using a complex and highly-trained model trained on low-level speech features provides unpredictable results, even when augmenting additional ``in-distribution'' training data; whereas higher-level speech features can help prevent failure.

    \subsection{Summary of Research Questions}
    \label{sec:experiments:summary}
    Through our experiments, we show that breaths are automatically and accurately detectable between the same speaker and throughout different speakers (\textbf{RQ1}), current in-the-wild deepfake speech does not produce breaths (\textbf{RQ2}), and using breaths as a discriminator between real and fake samples obtains performant results (\textbf{RQ3}).

\section{Discussion}
\label{sec:disc}
We discuss additional material that is important to the fundamental science of machine learning and deepfake discrimination.

\subsection{Reproducibility}
\label{sec:disc:repro}
Reproducibility is a growing concern, especially for machine learning~\cite{olszewski2023get}. To alleviate these issues and to aid future work and comparison, we make efforts to increase the reproducibility of our work. First, we publish the code and framework we use for our entire pipeline.\footnote{Code available upon publication.} Second, we publish the trained models and raw model scores files.\footnote{Scores file available upon publication.} We release the trained model and scores file as an additional artifact as it has been shown that retraining the same model on different GPUs may produce different results~\cite{abdullah2021sok}. This gives future researchers the ability to calculate \emph{any and all} metrics that may be desired for comparison.

\subsection{Limitations and Future Work}
\label{sec:disc:limitations}
If these deepfake samples start producing frequent breathing, our relatively simple discriminators will likely produce worse results. To accommodate this, a shift towards natural language processing (NLP) could be combined with our work. This combination would allow contextualization between the breaths that are identified and the intra-speech location relative to other parts of speech. This contextualization would minimize false positive breath detection and improve the results of a deepfake detector.


\section{Related Work}
\label{sec:rel_work}
  Recently, multiple papers have begun the exploration of breath importance in deepfake speech. However, none of these papers make any claims on \emph{current} deepfake speech deployed in the wild.  Mostaani et al. investigated whether breathing patterns are present in text-to-speech (TTS) and voice conversion (VC) algorithms using ASVspoof 2019~\cite{mostaani2022breathing}. This work showed that TTS algorithms fail to properly generate breaths while VC algorithms seem to retain breath pattern-related information. As TTS is rapidly dwarfing VC as the main proponent of deepfake speech, this work shows promise for breath usage as a speech deepfakes discriminator. Additionally, the Breathing-Talking-Silence Encoder (BTS-E) algorithm was proposed as an addition to existing countermeasures (CM) for voice spoofing attacks that use breath and silence event detection to enhance existing CM performance by up to 46\%~\cite{doan2023bts}. BTS-E leverages three Gaussian Mixture Models (GMM) to segment/label input audio into talking, silence, and breath segments. This segmentation is then translated into a latent feature space and combined with the last hidden layer in an existing CM to focus that CM on talking/breathing/silence events. However, the feature importance of breathing vs. silence is not explored in BTS-E; which is problematic as M\"{u}ller et al.~\cite{muller2021speech} identify an uneven distribution of leading and trailing silence duration in the ASVspoof 2019 and 2021 datasets, which correlate with the target prediction label. Thus, simply examining the silence patterns in the files allows for accurate classification of real/speech deepfakes. As such, it is unclear whether the breath features in BTS-E are important and the value of breaths is unknown. 

  Our work, instead, focuses solely on the prevalence of breaths and on samples gathered from real-world use. These real-world samples give an accurate measure of the landscape of deepfake speech as it stands currently. We make the first attempts at discriminating between real and fake in-the-wild speech using breathing.

\section{Conclusion}
\label{sec:conc}
    Deepfake speech generation advancements are reducing the gap between human-spoken and human-sounding audio. 
    Deepfake speech will become imperceivable to human speech, as such a focus needs to be placed on \emph{how} this speech is performed. Toward this, we employ a multi-tiered pipeline that focuses on breathing to discriminate between real/deepfake speech samples. We show that breaths are generalizable between speakers and that simple calculated breath features provide accurate classification results (perfect 1.0 AUPRC and 0.0 EER) on a dataset of in-the-wild real and synthetically-generated speech. Furthermore, we show the shortcomings of a complex deep learning model that fails to classify the same in-the-wild samples (0.72 AUPRC and 0.99 EER).


\bibliographystyle{ACM-Reference-Format}
\bibliography{ebydt}


\begin{thebibliography}{36}


\ifx \showCODEN    \undefined \def \showCODEN     #1{\unskip}     \fi
\ifx \showDOI      \undefined \def \showDOI       #1{#1}\fi
\ifx \showISBNx    \undefined \def \showISBNx     #1{\unskip}     \fi
\ifx \showISBNxiii \undefined \def \showISBNxiii  #1{\unskip}     \fi
\ifx \showISSN     \undefined \def \showISSN      #1{\unskip}     \fi
\ifx \showLCCN     \undefined \def \showLCCN      #1{\unskip}     \fi
\ifx \shownote     \undefined \def \shownote      #1{#1}          \fi
\ifx \showarticletitle \undefined \def \showarticletitle #1{#1}   \fi
\ifx \showURL      \undefined \def \showURL       {\relax}        \fi
\providecommand\bibfield[2]{#2}
\providecommand\bibinfo[2]{#2}
\providecommand\natexlab[1]{#1}
\providecommand\showeprint[2][]{arXiv:#2}

\bibitem[elm(2021)]%
        {elmers_take_2021}
 \bibinfo{year}{2021}\natexlab{}.
\newblock \bibinfo{booktitle}{\emph{Take a {Breath}: {Respiratory} {Sounds}
  {Improve} {Recollection} in {Synthetic} {Speech}}}.
\newblock
\urldef\tempurl%
\url{https://doi.org/10.21437/Interspeech.2021-1496}
\showDOI{\tempurl}
\newblock
\shownote{Pages: 3200}.


\bibitem[Abdullah et~al\mbox{.}(2021)]%
        {abdullah2021sok}
\bibfield{author}{\bibinfo{person}{Hadi Abdullah}, \bibinfo{person}{Kevin
  Warren}, \bibinfo{person}{Vincent Bindschaedler}, \bibinfo{person}{Nicolas
  Papernot}, {and} \bibinfo{person}{Patrick Traynor}.}
  \bibinfo{year}{2021}\natexlab{}.
\newblock \showarticletitle{Sok: The faults in our asrs: An overview of attacks
  against automatic speech recognition and speaker identification systems}. In
  \bibinfo{booktitle}{\emph{2021 IEEE symposium on security and privacy (SP)}}.
  IEEE, \bibinfo{pages}{730--747}.
\newblock


\bibitem[Amiriparian and Schuller({[n.\,d.]})]%
        {amiriparian2018deep}
\bibfield{author}{\bibinfo{person}{Shahin Amiriparian} {and}
  \bibinfo{person}{Bj{\"o}rn Schuller}.} \bibinfo{year}{[n.\,d.]}\natexlab{}.
\newblock \showarticletitle{Deep convolutional recurrent neural network for
  rare acoustic event detection}.
\newblock


\bibitem[Arp et~al\mbox{.}(2022)]%
        {arp2022and}
\bibfield{author}{\bibinfo{person}{Daniel Arp}, \bibinfo{person}{Erwin
  Quiring}, \bibinfo{person}{Feargus Pendlebury}, \bibinfo{person}{Alexander
  Warnecke}, \bibinfo{person}{Fabio Pierazzi}, \bibinfo{person}{Christian
  Wressnegger}, \bibinfo{person}{Lorenzo Cavallaro}, {and}
  \bibinfo{person}{Konrad Rieck}.} \bibinfo{year}{2022}\natexlab{}.
\newblock \showarticletitle{Dos and don'ts of machine learning in computer
  security}. In \bibinfo{booktitle}{\emph{31st USENIX Security Symposium
  (USENIX Security 22)}}. \bibinfo{pages}{3971--3988}.
\newblock


\bibitem[Attorresi et~al\mbox{.}(2022)]%
        {attorresi2022combining}
\bibfield{author}{\bibinfo{person}{Luigi Attorresi}, \bibinfo{person}{Davide
  Salvi}, \bibinfo{person}{Clara Borrelli}, \bibinfo{person}{Paolo Bestagini},
  {and} \bibinfo{person}{Stefano Tubaro}.} \bibinfo{year}{2022}\natexlab{}.
\newblock \showarticletitle{Combining automatic speaker verification and
  prosody analysis for synthetic speech detection}.
\newblock \bibinfo{journal}{\emph{arXiv preprint arXiv:2210.17222}}
  (\bibinfo{year}{2022}).
\newblock


\bibitem[Blue et~al\mbox{.}(2022)]%
        {blue2022you}
\bibfield{author}{\bibinfo{person}{Logan Blue}, \bibinfo{person}{Kevin Warren},
  \bibinfo{person}{Hadi Abdullah}, \bibinfo{person}{Cassidy Gibson},
  \bibinfo{person}{Luis Vargas}, \bibinfo{person}{Jessica O'Dell},
  \bibinfo{person}{Kevin Butler}, {and} \bibinfo{person}{Patrick Traynor}.}
  \bibinfo{year}{2022}\natexlab{}.
\newblock \showarticletitle{Who Are You (I Really Wanna Know)? Detecting Audio
  $\{$DeepFakes$\}$ Through Vocal Tract Reconstruction}. In
  \bibinfo{booktitle}{\emph{31st USENIX Security Symposium (USENIX Security
  22)}}. \bibinfo{pages}{2691--2708}.
\newblock


\bibitem[Braunschweiler and Chen(2013)]%
        {braunschweiler_automatic_2013}
\bibfield{author}{\bibinfo{person}{Norbert Braunschweiler} {and}
  \bibinfo{person}{Langzhou Chen}.} \bibinfo{year}{2013}\natexlab{}.
\newblock \showarticletitle{Automatic detection of inhalation breath pauses for
  improved pause modelling in {HMM}-{TTS}}.
\newblock \bibinfo{journal}{\emph{The ISCA Speech Synthesis Workshop}}
  (\bibinfo{year}{2013}), \bibinfo{pages}{6}.
\newblock


\bibitem[Br{\"u}mmer et~al\mbox{.}(2021)]%
        {brummer2021out}
\bibfield{author}{\bibinfo{person}{Niko Br{\"u}mmer}, \bibinfo{person}{Luciana
  Ferrer}, {and} \bibinfo{person}{Albert Swart}.}
  \bibinfo{year}{2021}\natexlab{}.
\newblock \showarticletitle{Out of a hundred trials, how many errors does your
  speaker verifier make?}
\newblock \bibinfo{journal}{\emph{arXiv preprint arXiv:2104.00732}}
  (\bibinfo{year}{2021}).
\newblock


\bibitem[Cvach(2012)]%
        {cvach2012monitor}
\bibfield{author}{\bibinfo{person}{Maria Cvach}.}
  \bibinfo{year}{2012}\natexlab{}.
\newblock \showarticletitle{Monitor alarm fatigue: an integrative review}.
\newblock \bibinfo{journal}{\emph{Biomedical instrumentation \& technology}}
  (\bibinfo{year}{2012}).
\newblock


\bibitem[Delgado et~al\mbox{.}(2021)]%
        {evalplan2021asvspoof}
\bibfield{author}{\bibinfo{person}{H{\'e}ctor Delgado},
  \bibinfo{person}{Nicholas Evans}, \bibinfo{person}{Tomi Kinnunen},
  \bibinfo{person}{Kong~Aik Lee}, \bibinfo{person}{Xuechen Liu},
  \bibinfo{person}{Andreas Nautsch}, \bibinfo{person}{Jose Patino},
  \bibinfo{person}{Md Sahidullah}, \bibinfo{person}{Massimiliano Todisco},
  \bibinfo{person}{Xin Wang}, {et~al\mbox{.}}} \bibinfo{year}{2021}\natexlab{}.
\newblock \showarticletitle{Asvspoof 2021: Automatic speaker verification
  spoofing and countermeasures challenge evaluation plan}.
\newblock \bibinfo{journal}{\emph{arXiv preprint arXiv:2109.00535}}
  (\bibinfo{year}{2021}).
\newblock


\bibitem[Deshpande and Schuller(2020)]%
        {deshpande_overview_2020}
\bibfield{author}{\bibinfo{person}{Gauri Deshpande} {and}
  \bibinfo{person}{Björn Schuller}.} \bibinfo{year}{2020}\natexlab{}.
\newblock \showarticletitle{An {Overview} on {Audio}, {Signal}, {Speech}, \&
  {Language} {Processing} for {COVID}-19}.
\newblock \bibinfo{journal}{\emph{arXiv:2005.08579 [cs, eess]}}
  (\bibinfo{date}{May} \bibinfo{year}{2020}).
\newblock
\newblock
\shownote{arXiv: 2005.08579}.


\bibitem[Doan et~al\mbox{.}(2023)]%
        {doan2023bts}
\bibfield{author}{\bibinfo{person}{Thien-Phuc Doan}, \bibinfo{person}{Long
  Nguyen-Vu}, \bibinfo{person}{Souhwan Jung}, {and} \bibinfo{person}{Kihun
  Hong}.} \bibinfo{year}{2023}\natexlab{}.
\newblock \showarticletitle{BTS-E: Audio Deepfake Detection Using
  Breathing-Talking-Silence Encoder}. In \bibinfo{booktitle}{\emph{ICASSP
  2023-2023 IEEE International Conference on Acoustics, Speech and Signal
  Processing (ICASSP)}}. IEEE, \bibinfo{pages}{1--5}.
\newblock


\bibitem[Fu et~al\mbox{.}(2019)]%
        {fu2019tuning}
\bibfield{author}{\bibinfo{person}{Guang-Hui Fu}, \bibinfo{person}{Lun-Zhao
  Yi}, {and} \bibinfo{person}{Jianxin Pan}.} \bibinfo{year}{2019}\natexlab{}.
\newblock \showarticletitle{Tuning model parameters in class-imbalanced
  learning with precision-recall curve}.
\newblock \bibinfo{journal}{\emph{Biometrical Journal}} \bibinfo{volume}{61},
  \bibinfo{number}{3} (\bibinfo{year}{2019}), \bibinfo{pages}{652--664}.
\newblock


\bibitem[Hamke et~al\mbox{.}({[n.\,d.]})]%
        {hamke_breath_nodate}
\bibfield{author}{\bibinfo{person}{Eric~E Hamke}, \bibinfo{person}{Ramiro
  Jordan}, {and} \bibinfo{person}{Manel Ramon-Martinez}.}
  \bibinfo{year}{[n.\,d.]}\natexlab{}.
\newblock \showarticletitle{Breath {Activity} {Detection} {Algorithm}}.
\newblock  (\bibinfo{year}{[n.\,d.]}), \bibinfo{pages}{11}.
\newblock


\bibitem[Hosler et~al\mbox{.}(2021)]%
        {hosler2021deepfakes}
\bibfield{author}{\bibinfo{person}{Brian Hosler}, \bibinfo{person}{Davide
  Salvi}, \bibinfo{person}{Anthony Murray}, \bibinfo{person}{Fabio Antonacci},
  \bibinfo{person}{Paolo Bestagini}, \bibinfo{person}{Stefano Tubaro}, {and}
  \bibinfo{person}{Matthew~C Stamm}.} \bibinfo{year}{2021}\natexlab{}.
\newblock \showarticletitle{Do deepfakes feel emotions? A semantic approach to
  detecting deepfakes via emotional inconsistencies}. In
  \bibinfo{booktitle}{\emph{Proceedings of the IEEE/CVF conference on computer
  vision and pattern recognition}}. \bibinfo{pages}{1013--1022}.
\newblock


\bibitem[Jung et~al\mbox{.}(2022)]%
        {jung2022sasv}
\bibfield{author}{\bibinfo{person}{Jee-weon Jung}, \bibinfo{person}{Hemlata
  Tak}, \bibinfo{person}{Hye-jin Shim}, \bibinfo{person}{Hee-Soo Heo},
  \bibinfo{person}{Bong-Jin Lee}, \bibinfo{person}{Soo-Whan Chung},
  \bibinfo{person}{Ha-Jin Yu}, \bibinfo{person}{Nicholas Evans}, {and}
  \bibinfo{person}{Tomi Kinnunen}.} \bibinfo{year}{2022}\natexlab{}.
\newblock \showarticletitle{SASV 2022: The first spoofing-aware speaker
  verification challenge}.
\newblock \bibinfo{journal}{\emph{arXiv preprint arXiv:2203.14732}}
  (\bibinfo{year}{2022}).
\newblock


\bibitem[Kinnunen et~al\mbox{.}(2017)]%
        {kinnunen2017asvspoof}
\bibfield{author}{\bibinfo{person}{Tomi Kinnunen}, \bibinfo{person}{Md
  Sahidullah}, \bibinfo{person}{H{\'e}ctor Delgado},
  \bibinfo{person}{Massimiliano Todisco}, \bibinfo{person}{Nicholas Evans},
  \bibinfo{person}{Junichi Yamagishi}, {and} \bibinfo{person}{Kong~Aik Lee}.}
  \bibinfo{year}{2017}\natexlab{}.
\newblock \showarticletitle{The ASVspoof 2017 challenge: Assessing the limits
  of replay spoofing attack detection}.
\newblock  (\bibinfo{year}{2017}).
\newblock


\bibitem[Layton et~al\mbox{.}(2024)]%
        {needlestack_layton}
\bibfield{author}{\bibinfo{person}{Seth Layton}, \bibinfo{person}{Tyler
  Tucker}, \bibinfo{person}{Daniel Olszewski}, \bibinfo{person}{Kevin Warren},
  \bibinfo{person}{Kevin Butler}, {and} \bibinfo{person}{Patrick Traynor}.}
  \bibinfo{year}{2024}\natexlab{}.
\newblock \showarticletitle{{SoK: The Good, The Bad, and The Unbalanced:
  Measuring Structural Limitations of Deepfake Datasets}}. In
  \bibinfo{booktitle}{\emph{{Proceedings of the USENIX Security Symposium
  (Security)}}}.
\newblock


\bibitem[Lorenzo-Trueba et~al\mbox{.}(2018)]%
        {lorenzo-trueba_can_2018}
\bibfield{author}{\bibinfo{person}{Jaime Lorenzo-Trueba},
  \bibinfo{person}{Fuming Fang}, \bibinfo{person}{Xin Wang},
  \bibinfo{person}{Isao Echizen}, \bibinfo{person}{Junichi Yamagishi}, {and}
  \bibinfo{person}{Tomi Kinnunen}.} \bibinfo{year}{2018}\natexlab{}.
\newblock \showarticletitle{Can we steal your vocal identity from the
  Internet?: Initial investigation of cloning Obama’s voice using GAN,
  WaveNet and low-quality found data}. In \bibinfo{booktitle}{\emph{Speaker and
  Language Recognition Workshop}}.
\newblock


\bibitem[Mostaani and Doss(2022)]%
        {mostaani2022breathing}
\bibfield{author}{\bibinfo{person}{Zohreh Mostaani} {and}
  \bibinfo{person}{Mathew~Magimai Doss}.} \bibinfo{year}{2022}\natexlab{}.
\newblock \showarticletitle{On breathing pattern information in synthetic
  speech}. In \bibinfo{booktitle}{\emph{Proc. Interspeech}}.
\newblock


\bibitem[M{\"u}ller et~al\mbox{.}(2021)]%
        {muller2021speech}
\bibfield{author}{\bibinfo{person}{Nicolas~M M{\"u}ller},
  \bibinfo{person}{Franziska Dieckmann}, \bibinfo{person}{Pavel Czempin},
  \bibinfo{person}{Roman Canals}, \bibinfo{person}{Konstantin B{\"o}ttinger},
  {and} \bibinfo{person}{Jennifer Williams}.} \bibinfo{year}{2021}\natexlab{}.
\newblock \showarticletitle{Speech is silver, silence is golden: What do
  ASVspoof-trained models really learn?}
\newblock \bibinfo{journal}{\emph{arXiv preprint arXiv:2106.12914}}
  (\bibinfo{year}{2021}).
\newblock


\bibitem[Nallanthighal et~al\mbox{.}(2021)]%
        {nallanthighal_deep_2021}
\bibfield{author}{\bibinfo{person}{Venkata~Srikanth Nallanthighal},
  \bibinfo{person}{Zohreh Mostaani}, \bibinfo{person}{Aki Härmä},
  \bibinfo{person}{Helmer Strik}, {and} \bibinfo{person}{Mathew Magimai-Doss}.}
  \bibinfo{year}{2021}\natexlab{}.
\newblock \showarticletitle{Deep learning architectures for estimating
  breathing signal and respiratory parameters from speech recordings}.
\newblock \bibinfo{journal}{\emph{Neural Networks}}  \bibinfo{volume}{141}
  (\bibinfo{year}{2021}), \bibinfo{pages}{211--224}.
\newblock
\showISSN{0893-6080}


\bibitem[Nallanthighal and Strik(2019)]%
        {nallanthighal2019deep}
\bibfield{author}{\bibinfo{person}{Venkata~Srikanth Nallanthighal} {and}
  \bibinfo{person}{H Strik}.} \bibinfo{year}{2019}\natexlab{}.
\newblock \showarticletitle{Deep sensing of breathing signal during
  conversational speech}.
\newblock  (\bibinfo{year}{2019}).
\newblock


\bibitem[Olszewski et~al\mbox{.}(2023)]%
        {olszewski2023get}
\bibfield{author}{\bibinfo{person}{Daniel Olszewski}, \bibinfo{person}{Allison
  Lu}, \bibinfo{person}{Carson Stillman}, \bibinfo{person}{Kevin Warren},
  \bibinfo{person}{Cole Kitroser}, \bibinfo{person}{Alejandro Pascual},
  \bibinfo{person}{Divyajyoti Ukirde}, \bibinfo{person}{Kevin Butler}, {and}
  \bibinfo{person}{Patrick Traynor}.} \bibinfo{year}{2023}\natexlab{}.
\newblock \showarticletitle{"Get in Researchers; We're Measuring
  Reproducibility": A Reproducibility Study of Machine Learning Papers in Tier
  1 Security Conferences}. In \bibinfo{booktitle}{\emph{Proceedings of the 2023
  ACM SIGSAC Conference on Computer and Communications Security}}.
  \bibinfo{pages}{3433--3459}.
\newblock


\bibitem[Paris and Donovan(2019)]%
        {paris2019deepfakes}
\bibfield{author}{\bibinfo{person}{Britt Paris} {and} \bibinfo{person}{Joan
  Donovan}.} \bibinfo{year}{2019}\natexlab{}.
\newblock \showarticletitle{Deepfakes and cheap fakes}.
\newblock  (\bibinfo{year}{2019}).
\newblock


\bibitem[Saunders(2019)]%
        {SaundersDetectingDF}
\bibfield{author}{\bibinfo{person}{Jonathan Saunders}.}
  \bibinfo{year}{2019}\natexlab{}.
\newblock \showarticletitle{Detecting Deep Fakes With Mice : Machines vs
  Biology}. In \bibinfo{booktitle}{\emph{Black Hat USA}}.
\newblock


\bibitem[Schuller et~al\mbox{.}(2020)]%
        {schuller_interspeech_2020}
\bibfield{author}{\bibinfo{person}{Björn~W. Schuller}, \bibinfo{person}{Anton
  Batliner}, \bibinfo{person}{Christian Bergler}, \bibinfo{person}{Eva-Maria
  Messner}, \bibinfo{person}{Antonia Hamilton}, \bibinfo{person}{Shahin
  Amiriparian}, \bibinfo{person}{Alice Baird}, \bibinfo{person}{Georgios
  Rizos}, \bibinfo{person}{Maximilian Schmitt}, \bibinfo{person}{Lukas
  Stappen}, \bibinfo{person}{Harald Baumeister},
  \bibinfo{person}{Alexis~Deighton MacIntyre}, {and} \bibinfo{person}{Simone
  Hantke}.} \bibinfo{year}{2020}\natexlab{}.
\newblock \showarticletitle{The {INTERSPEECH} 2020 {Computational}
  {Paralinguistics} {Challenge}: {Elderly} {Emotion}, {Breathing} \& {Masks}}.
  In \bibinfo{booktitle}{\emph{Interspeech 2020}}. \bibinfo{publisher}{ISCA},
  \bibinfo{pages}{2042--2046}.
\newblock


\bibitem[Sugrim et~al\mbox{.}(2019)]%
        {sugrim2019robust}
\bibfield{author}{\bibinfo{person}{Shridatt Sugrim}, \bibinfo{person}{Can Liu},
  \bibinfo{person}{Meghan McLean}, {and} \bibinfo{person}{Janne Lindqvist}.}
  \bibinfo{year}{2019}\natexlab{}.
\newblock \showarticletitle{Robust performance metrics for authentication
  systems}. In \bibinfo{booktitle}{\emph{Network and Distributed Systems
  Security (NDSS) Symposium 2019}}.
\newblock


\bibitem[Sz{\'e}kely et~al\mbox{.}(2019)]%
        {szekely2019casting}
\bibfield{author}{\bibinfo{person}{{\'E}va Sz{\'e}kely},
  \bibinfo{person}{Gustav~Eje Henter}, {and} \bibinfo{person}{Joakim
  Gustafson}.} \bibinfo{year}{2019}\natexlab{}.
\newblock \showarticletitle{Casting to corpus: Segmenting and selecting
  spontaneous dialogue for TTS with a CNN-LSTM speaker-dependent breath
  detector}. In \bibinfo{booktitle}{\emph{ICASSP 2019-2019 IEEE International
  Conference on Acoustics, Speech and Signal Processing (ICASSP)}}. IEEE,
  \bibinfo{pages}{6925--6929}.
\newblock


\bibitem[Székely et~al\mbox{.}(2020)]%
        {szekely_breathing_2020}
\bibfield{author}{\bibinfo{person}{Éva Székely}, \bibinfo{person}{Gustav~Eje
  Henter}, \bibinfo{person}{Jonas Beskow}, {and} \bibinfo{person}{Joakim
  Gustafson}.} \bibinfo{year}{2020}\natexlab{}.
\newblock \showarticletitle{Breathing and {Speech} {Planning} in {Spontaneous}
  {Speech} {Synthesis}}. In \bibinfo{booktitle}{\emph{{ICASSP} 2020 - 2020
  {IEEE} {International} {Conference} on {Acoustics}, {Speech} and {Signal}
  {Processing} ({ICASSP})}}. \bibinfo{pages}{7649--7653}.
\newblock
\newblock
\shownote{ISSN: 2379-190X}.


\bibitem[Tak et~al\mbox{.}(2022)]%
        {tak2022automatic}
\bibfield{author}{\bibinfo{person}{Hemlata Tak}, \bibinfo{person}{Massimiliano
  Todisco}, \bibinfo{person}{Xin Wang}, \bibinfo{person}{Jee-weon Jung},
  \bibinfo{person}{Junichi Yamagishi}, {and} \bibinfo{person}{Nicholas Evans}.}
  \bibinfo{year}{2022}\natexlab{}.
\newblock \showarticletitle{Automatic speaker verification spoofing and
  deepfake detection using wav2vec 2.0 and data augmentation}.
\newblock \bibinfo{journal}{\emph{arXiv preprint arXiv:2202.12233}}
  (\bibinfo{year}{2022}).
\newblock


\bibitem[Todisco et~al\mbox{.}(2019)]%
        {todisco2019asvspoof}
\bibfield{author}{\bibinfo{person}{Massimiliano Todisco}, \bibinfo{person}{Xin
  Wang}, \bibinfo{person}{Ville Vestman}, \bibinfo{person}{Md Sahidullah},
  \bibinfo{person}{H{\'e}ctor Delgado}, \bibinfo{person}{Andreas Nautsch},
  \bibinfo{person}{Junichi Yamagishi}, \bibinfo{person}{Nicholas Evans},
  \bibinfo{person}{Tomi Kinnunen}, {and} \bibinfo{person}{Kong~Aik Lee}.}
  \bibinfo{year}{2019}\natexlab{}.
\newblock \showarticletitle{ASVspoof 2019: Future horizons in spoofed and fake
  audio detection}.
\newblock \bibinfo{journal}{\emph{arXiv preprint arXiv:1904.05441}}
  (\bibinfo{year}{2019}).
\newblock


\bibitem[Warren et~al\mbox{.}(2024)]%
        {WTC_24}
\bibfield{author}{\bibinfo{person}{Kevin Warren}, \bibinfo{person}{Tyler
  Tucker}, \bibinfo{person}{Anna Crowder}, \bibinfo{person}{Daniel Olszewski},
  \bibinfo{person}{Allison Lu}, \bibinfo{person}{Caroline Fedele},
  \bibinfo{person}{Magdalena Pasternak}, \bibinfo{person}{Seth Layton},
  \bibinfo{person}{Kevin Butler}, \bibinfo{person}{Carrie Gates}, {and}
  \bibinfo{person}{Patrick Traynor}.} \bibinfo{year}{2024}\natexlab{}.
\newblock \showarticletitle{{Better Be Computer or I'm Dumb": A Large-Scale
  Evaluation of Humans as Audio Deepfake Detectors}}. In
  \bibinfo{booktitle}{\emph{{Proceedings of the ACM Conference on Computer and
  Communications Security (CCS)}}}.
\newblock


\bibitem[Wu et~al\mbox{.}(2015)]%
        {wu2015asvspoof}
\bibfield{author}{\bibinfo{person}{Zhizheng Wu}, \bibinfo{person}{Tomi
  Kinnunen}, \bibinfo{person}{Nicholas Evans}, \bibinfo{person}{Junichi
  Yamagishi}, \bibinfo{person}{Cemal Hanil{\c{c}}i}, \bibinfo{person}{Md
  Sahidullah}, {and} \bibinfo{person}{Aleksandr Sizov}.}
  \bibinfo{year}{2015}\natexlab{}.
\newblock \showarticletitle{ASVspoof 2015: the first automatic speaker
  verification spoofing and countermeasures challenge}. In
  \bibinfo{booktitle}{\emph{Sixteenth annual conference of the international
  speech communication association}}.
\newblock


\bibitem[Yamagishi et~al\mbox{.}(2021)]%
        {yamagishi2021asvspoof}
\bibfield{author}{\bibinfo{person}{Junichi Yamagishi}, \bibinfo{person}{Xin
  Wang}, \bibinfo{person}{Massimiliano Todisco}, \bibinfo{person}{Md
  Sahidullah}, \bibinfo{person}{Jose Patino}, \bibinfo{person}{Andreas
  Nautsch}, \bibinfo{person}{Xuechen Liu}, \bibinfo{person}{Kong~Aik Lee},
  \bibinfo{person}{Tomi Kinnunen}, \bibinfo{person}{Nicholas Evans},
  {et~al\mbox{.}}} \bibinfo{year}{2021}\natexlab{}.
\newblock \showarticletitle{ASVspoof 2021: accelerating progress in spoofed and
  deepfake speech detection}.
\newblock \bibinfo{journal}{\emph{arXiv preprint arXiv:2109.00537}}
  (\bibinfo{year}{2021}).
\newblock


\bibitem[Yi et~al\mbox{.}(2022)]%
        {yi2022add}
\bibfield{author}{\bibinfo{person}{Jiangyan Yi}, \bibinfo{person}{Ruibo Fu},
  \bibinfo{person}{Jianhua Tao}, \bibinfo{person}{Shuai Nie},
  \bibinfo{person}{Haoxin Ma}, \bibinfo{person}{Chenglong Wang},
  \bibinfo{person}{Tao Wang}, \bibinfo{person}{Zhengkun Tian},
  \bibinfo{person}{Ye Bai}, \bibinfo{person}{Cunhang Fan}, {et~al\mbox{.}}}
  \bibinfo{year}{2022}\natexlab{}.
\newblock \showarticletitle{Add 2022: the first audio deep synthesis detection
  challenge}. In \bibinfo{booktitle}{\emph{ICASSP 2022-2022 IEEE International
  Conference on Acoustics, Speech and Signal Processing (ICASSP)}}. IEEE,
  \bibinfo{pages}{9216--9220}.
\newblock


\end{thebibliography}

\end{document}